\documentstyle [12pt,a4p,epsfig,amsmath,multicol]{article}
\begin{document}
\vspace*{0.6cm}

\begin{center} 
{\normalsize\bf A Covariant Feynman Path Amplitude Calculation
  of Neutrino and Muon Oscillations}
\end{center}
\vspace*{0.6cm}
\centerline{\footnotesize J.H.Field}
\baselineskip=13pt
\centerline{\footnotesize\it D\'{e}partement de Physique Nucl\'{e}aire et 
 Corpusculaire, Universit\'{e} de Gen\`{e}ve}
\baselineskip=12pt
\centerline{\footnotesize\it 24, quai Ernest-Ansermet CH-1211Gen\`{e}ve 4. }
\centerline{\footnotesize E-mail: john.field@cern.ch}
\baselineskip=13pt
 
\vspace*{0.9cm}
\abstract{ Predictions are presented for the oscillation phase in neutrino
  oscillations following pion decay at rest and in flight, muon decay and 
  nuclear $\beta-$decay at rest as well as for muon oscillations
  following pion decay at rest and in flight. The neutrino oscillation
  phases found disagree with the conventionally used value: $\phi_{12} = \Delta m^2 L/(2P)$.
  The same oscillation phase is found for neutrino and muon oscillations following 
   pion decay at rest.
 Shortcomings of previous treatments of the quantum mechanics of neutrino oscillations
 are commented on.}
\vspace*{0.9cm}
\normalsize\baselineskip=15pt
\setcounter{footnote}{0}
\renewcommand{\thefootnote}{\alph{footnote}}
\newline
{\it Keywords ;} Quantum Mechanics,
Neutrino Oscillations.
\newline

\vspace*{0.4cm}

 This letter sketches briefly a derivation of spatially dependent interference
 effects in the detection probability of neutrinos and muons following the
  weak decays of unstable `source' particles: pions, muons and $\beta-$radioactive
 nuclei. It is based on Feynman's path amplitude formulation of quantum
 mechanics~\cite{Feyn2} in which the probability of transition from a
 set of initial states $I = \sum_l i_l$ to a set of final states $F = \sum_m f_m$
 is given by the relation:
  \begin{equation}
 P_{FI} = \sum_m \sum_l \left|\sum_{k_1} \sum_{k_2}...\sum_{k_n}\langle f_m| k_1 \rangle
\langle k_1| k_2 \rangle...\langle k_n| i_l \rangle \right|^2  
 \end{equation}
 where  $k_j,~j=1,n$ are (unobserved) intermediate quantum states.
 Only the essential elements of the calculation and the results are presented
 here. Full details are given elsewhere~\cite{JHF}.
 \par The application of 
  Eqn(1) to pion decay at rest is illustrated in the ideal experiment shown in
  Fig.1.  A $\pi^+$ comes to rest in a stopping target T at time $t_0$ as recorded
  by the counter C$_{\rm A}$ (Fig.1a)). The pion at rest constitutes the initial state
  of the path amplitudes.
  In Fig.1b) and Fig.1c) are shown two alternative histories of the stopped pion.
  In Fig.1b) the pion decays at time $t_1$ into the neutrino mass eigenstate
  $|\nu_1>$, of mass $m_1$, and in Fig.1c) into the neutrino mass eigenstate
  $|\nu_2>$, of mass $m_2$, at the later time $t_2$.  If $m_1>m_2$, then, for a 
  suitable choice of the times $t_1$ and $t_2$, interference between the path amplitudes
  corresponding to the different physical processes shown in  Fig.1b) and Fig.1c)
  will occur when a neutrino interaction $\nu_e n \rightarrow e^- p$ takes place at time
  $t_D$ and distance $L$ from the pion decay point(Fig.1d)). The final state of the neutrino
  interaction event is also that of the path amplitudes.
  For this experiment, the path amplitudes corresponding to the
  two alternative histories of the decaying pion are:
\begin{eqnarray}
 A_i&=& <e^-p|T|n \nu_e><\nu_e|\nu_i>D( x_f-x_i,t_D-t_i, m_i )<\nu_i|\nu_\mu>  \nonumber \\ 
 &  & <\nu_\mu \mu^+|T|\pi^+> e^{-\frac{\Gamma_\pi}{2}(t_i-t_0)}D( 0,t_i-t_0, m_\pi )~~i=1,2 
  \end{eqnarray} 
 Here   $<e^-p|T|n \nu_e>$ and $<\nu_\mu \mu^+|T|\pi^+>$ are the invariant amplitudes of the
  neutrino interaction and pion decay processes respectively, $<\nu_e|\nu_i>$ and
  $<\nu_i|\nu_\mu>$ are amplitudes describing the mixture of the flavour and mass
  neutrino eigenstates:
\begin{eqnarray}
|\nu_e>& = &\cos\theta|\nu_1>+\sin\theta|\nu_2> \\
|\nu_\mu>& = &-\sin\theta|\nu_1>+\cos\theta|\nu_2> 
\end{eqnarray}
(only 2-flavour mixing is considered), and $D$ is the Lorentz invariant
 configuration space propagator~\cite{Feyn1,Moh1} of a neutrino or the pion. In the limit
 of large time-like separations or of on-shell particles, appropriate
 to the experiment shown in Fig.1, 
 $D \simeq \exp[-im\Delta \tau]$ where $m$ is the pole mass of the
   particle and $\Delta \tau$ the increment of proper time corresponding to the 
  path. In the following the additional functional dependence $\simeq (m/\Delta \tau)^\frac{3}{2}$
  of $D$ in the asymptotic region (leading to solid angle correction factors) is neglected.
  With these approximations:
 \begin{eqnarray}
 D(\Delta x,\Delta t,m) & \simeq &  \exp[-im \sqrt{(\Delta t)^2-(\Delta x)^2}] \nonumber \\
      & = &  \exp[-i m \Delta \tau] \nonumber \\  
      & \equiv & \exp[-i\Delta \phi] 
 \end{eqnarray}
 The phase increments, $\Delta \phi$, corresponding to the paths of the neutrinos and
 the pion in the amplitudes $A_i$ are:
\begin{eqnarray}
\Delta \phi_i^{\nu}& = & m_i \Delta \tau_i =\frac{m_i^2}{E_i}
 \Delta t_i = \frac{m_i^2}{P_i}L   \\
 \Delta \phi_i^{\pi}& = & m_{\pi} (t_i-t_0)
  =  m_{\pi} (t_D-t_0)-\frac{m_{\pi}L}{v_i}  \nonumber \\
 & = & m_{\pi} (t_D-t_0)-m_{\pi}L\left\{1+\frac{m_i^2}{2 P_0^2}\right\}
\end{eqnarray}
  where $m_i$, $E_i$, $P_i$ and $v_i$ are the mass, energy, momentum and velocity
  of the mass eigenstate $\nu_i$ and
 \begin{equation}
 P_0 = \frac{m_{\pi}^2-m_{\mu}^2}{2 m_{\pi}}~ = ~ 29.8 \rm{MeV}
\end{equation}
  The neutrinos are assumed to follow classical rectilinear trajectories
 such that $\Delta t_i = L/ v_i = E_iL/P_i$ and the time dilatation formula
 $\Delta t =  \gamma \Delta \tau = E \Delta \tau/m$ has been used in Eqn(6).
 In Eqn(7) the neutrino velocity\footnote{Units with $h/2\pi = c = 1$ are used.}
 $v_i$ is expressed in terms of the neutrino mass to order $m_i^2$.
 Using Eqns(5-7) the path amplitudes of Eqn(2) may be written as:
 \begin{eqnarray}
 A_i & = & <e^-p|T|n \nu_e><\nu_e|\nu_i><\nu_i|\nu_\mu><\nu_\mu \mu^+|T|\pi^+> \nonumber \\
 &   &  
 \exp[ i\phi_0-\frac{\Gamma_\pi}{2}(t_D-t_0-t_i^{fl})] 
\exp i \left[ \frac{ m_i^2}{P_0}\left(
 \frac{m_{\pi}}{2 P_0}-1\right)L \right]~~~i=1,2 
 \end{eqnarray} 
 where the neutrino times-of-flight $t_i^{fl} = t_D - t_i$ have been introduced
 and 
\begin{equation}
\phi_0 \equiv m_{\pi}(L-t_D+t_0)
\end{equation}
 Using now Eqn(1) to calculate the transition probability, and integrating over
 the detection time $t_D$~\cite{JHF}, gives, for the probability to observe the
 reaction $\nu_en \rightarrow e^- p$ at distance $L$ from the decay point:
\begin{equation}
 P(e^-p|L) = C_N(\nu;\pi)\ \sin^2 \theta \cos^2 \theta(1-F^{\nu}(\Gamma_{\pi})\cos \phi_{12}^{\nu,\pi})
\end{equation}
where 
\begin{equation}
\phi_{12}^{\nu,\pi} = \frac{\Delta m^2}{P_0}\left(
 \frac{m_{\pi}}{2 P_0}-1\right)L = \frac{2 m_{\pi} m_{\mu}^2 \Delta m^2 L}
 {(m_{\pi}^2-m_{\mu}^2)^2} 
\end{equation}
\begin{equation}
F^{\nu}(\Gamma_{\pi})=\exp\left(-\frac{\Gamma_{\pi} m_{\pi}}{2 m_{\mu}^2}
 \phi_{12}^{\nu,\pi}\right) 
\end{equation}
 and Eqns(3) and (4) have been used. Here $C_N(\nu;\pi)$ is an $L$ independent
 normalisation factor and $\Delta m^2 \equiv m_1^2-m_2^2$. The first
 and second terms in the second member of Eqn(12) are the contributions
 of $\Delta \phi_i^{\pi}$ and  $\Delta \phi_i^{\nu}$ to the interference
 phase. The latter is a factor of two larger than in the conventional
 value~\cite{BilPont2, BiPet} $\phi_{12} = \Delta m^2 L/(2P)$.
    $\Delta \phi_i^{\pi}$ gives a numerically large
 ($m_{\pi}/2P_0=2.34$) contribution to the oscillation phase.   
 \par The above calculation is readily repeated for the case of 
 detection of muon decay $\mu^+ \rightarrow e^+ \nu_e \overline{\nu}_{\mu}$
 at distance $L$ from the $\pi^+$ decay point. The phase increments
 analagous to Eqns(6) and (7) are:
\begin{eqnarray} 
\Delta \phi^{\mu}_i & = & \frac{m_{\mu}^2 L }{P_0}\left[
 1+\frac{m_i^2 E_0^{\mu}}{2 m_{\pi} P_0^2}\right]  \\
 \Delta \phi_i^{\pi(\mu)} & = & m_{\pi} (t_D-t_0)
 -\frac{m_{\pi}L}{v^{\mu}_0}\left[1+\frac{4 m_i^2 m_{\pi}^2 m_{\mu}^2}
{(m_{\pi}^2-m_{\mu}^2)^2(m_{\pi}^2+m_{\mu}^2)} \right]  
\end{eqnarray}    
where 
\[ E_0^{\mu} = \frac{m_{\pi}^2+m_{\mu}^2}{2 m_{\pi}}~~~~~~~{\rm and} 
 ~~~~~~~ 
v^{\mu}_0 = \frac{m_{\pi}^2-m_{\mu}^2}{m_{\pi}^2+m_{\mu}^2} \]
 The result found for the time-integrated decay probability is:
\begin{equation}
 P(e^+\nu_e \overline{\nu}_{\mu} |L) = C_N(\mu;\pi)(1-F^{\mu}(\Gamma_{\pi})
\sin 2 \theta \cos \phi_{12}^{\mu,\pi})
\end{equation}
where 
\begin{equation}
 \phi_{12}^{\mu,\pi} =  \frac{m_{\mu}^2 \Delta m^2}{2 P_0^3}
\left(1-\frac{ E_0^{\mu}}{m_{\pi}}\right)L = \frac{2 m_{\pi} m_{\mu}^2 \Delta m^2 L}
 {(m_{\pi}^2-m_{\mu}^2)^2} = \phi_{12}^{\nu,\pi}
\end{equation}
\begin{equation}
F^{\mu}(\Gamma_{\pi})= \exp \left(-\frac{\Gamma_{\pi}
 m_{\pi}}{(m_{\pi}^2-m_{\pi}^2)} \phi_{12}^{\mu,\pi} \right)
\end{equation}
The first
 and second terms in the second member of Eqn(17) are the contributions
 of $\Delta \phi_i^{\pi(\mu)}$ and  $\Delta \phi_i^{\mu}$ to the interference
 phase. It is interesting to note that neutrino and muon oscillation
 phases are the same for given values of $\Delta m^2$ and $L$.
 For oscillation phases $\phi_{12}^{\nu,\pi} = \phi_{12}^{\mu,\pi} =1$, the damping
  factors of the oscillation term, due to the non-vanishing pion lifetime
  take the values $F^{\nu}(\Gamma_{\pi}) = 1-1.58\times10^{-16}$ and 
$F^{\mu}(\Gamma_{\pi}) = 1-4.4\times10^{-16}$. This damping effect is thus 
  completely negligible in typical neutrino oscillation experiments with
  oscillation phases of order unity. 
\par Formulae like Eqn(11) have been derived in a similar manner~\cite{JHF}
 for neutrino oscillations
 $\overline{\nu}_{\mu} \rightarrow \overline{\nu}_e$ following muon decay at rest, 
 detected  via the process $ \overline{\nu}_e p \rightarrow e^+ n$ and 
  $\overline{\nu}_e \rightarrow \overline{\nu}_e$ oscillations following
  nuclear $\beta-$decay, detected via
 the process $\overline{\nu}_e p \rightarrow e^+n$. 
  The results found, for the time-integrated decay probabilities, are:
\begin{eqnarray}
 P(e^+n,\mu|L) & = & C_N(\overline{\nu};\mu) \sin^2 \theta \cos^2 \theta\left[1-\cos
 \frac{\Delta m^2}{P_{\overline{\nu}}}\left(
 \frac{m_{\mu}}{2 P_{\overline{\nu}}}-1\right)L\right]  \\ 
 P(e^+n,\beta|L) & = & C_N(\overline{\nu};\beta)
 \left[ \sin^4\theta + \cos^4\theta+2 \sin^2 \theta \cos^2
 \theta \cos \frac{\Delta m^2}{P_{\overline{\nu}}}\left(
 \frac{E_{\beta}}{2 P_{\overline{\nu}}}-1 \right)L \right]~ 
\end{eqnarray}
In these formulae, $ P_{\overline{\nu}}$ is the momentum of the detected
 $\overline{\nu}_e$, $E_{\beta}$ is the total energy release in the
 $\beta-$decay process, and damping corrections due to the finite lifetimes of
  the decaying particles are neglected.
\par Finally, in Ref.~\cite{JHF}, the cases of neutrino and muon oscillations
 following the decay in flight of ultrarelativistic pions was considered.
 The oscillation probability formulae are the same as Eqns(11) and (16)
 respectively, with the phases:
\begin{eqnarray}
\phi_{12}^{\nu,\pi}(\rm{in~flight}) & = & \frac{ m_{\mu}^2 \Delta m^2 L}
{(m_{\pi}^2-m_{\mu}^2) E_{\nu} \cos \theta_{\nu}}   \\
 \phi_{12}^{\mu,\pi}(\rm{in~flight}) & = & \frac{2 m_{\mu}^2 \Delta m^2 ( m_{\mu}^2 E_{\pi}-
m_{\pi}^2 E_{\mu})L}{(m_{\pi}^2-m_{\mu}^2)^2 E_{\mu}^2 \cos \theta_{\mu} }
\end{eqnarray}
 The neutrinos or muons are detected at distance $L$ from the decay point,
 along the direction of flight of the parent pion, with decay angles $\theta_{\nu}$
 and $\theta_{\mu}$ respectively.
 \par Because the parent pion and the daughter muon in the decay process 
 $\pi \rightarrow  \mu \nu$ are unstable particles, their physical masses
 $W_{\pi}$ and $W_{\mu}$  have distributions depending, through Breit-Wigner
 amplitudes, on their pole masses $m_{\pi}$ and $m_{\mu}$ and decay widths
 $\Gamma_{\pi}$ and $\Gamma_{\mu}$. Energy-momentum conservation in the 
  pion decay process then leads to a distribution of path amplitudes 
  with different neutrino momenta, an effect neglected in the above
  discussion. In Ref.\cite{JHF} corrections resulting from the (coherent)
 momentum smearing due to $W_{\mu}$ and (incoherent) smearing due to
 $W_{\pi}$ are calculated using a Gaussian approximation for
 the Breit-Wigner amplitudes. The resulting damping corrections
  due to $W_{\pi}$ to the
  interference terms in Eqns(11) and (16) are found to be vanishingly small.
  For $\Delta m^2 =$ (1eV)$^2$ and $L = 30$m (typical of the LNSD~\cite{LNSD}
  or KARMEN~\cite{KARMEN} experiments) the corresponding damping factors
  are found to be $1-1.3\times10^{-29}$ (for neutrino oscillations)
  and $1-3.9 \times 10^{-31}$ (for muon oscillations)~\cite{JHF}.
  The damping corrections from the variation of $W_{\mu}$ are even smaller.
  \par Corrections to Eqns(11) and (16) due to thermal motion of the decaying 
  pion and finite target and detector sizes have also been evaluated in
  Ref.\cite{JHF}. For neutrino oscillations with $\Delta m^2 =$ (1eV)$^2$
  and $L = 30$m and a room-temperature target, the damping factor of the
  interference term is found to be $1-6.7\times10^{-10}$ and the shift
  in the oscillation phase to be $1.2\times10^{-9}$rad. In summary, all
  known sources of damping of the interference terms in Eqn(10) and (15)
  are expected to be completely negligible in any forseeable neutrino
  or muon oscillation experiment.
   \par The first published calculation of the neutrino oscillation
   phase~\cite{GribPont} gave a result (only the contribution of the neutrino
   propagators was taken into account) in agreement with Eqn(6) above.
  A later calculation~\cite{FritMink} assumed instead that the phase of
  the neutrino propagator evolves as $Et$, i.e. according to the
  non-relativistic Schr\"{o}dinger equation. The assumption was 
  also made that the different neutrino mass states have equal momenta. 
  This leads to the `standard' formula for the oscillation phase:
   $\phi_{12} = \Delta m^2 L/(2P)$~\cite{BilPont2, BiPet}
   which has subsequently been
   used for the analysis of all neutrino oscillation experiments.
   The Lorentz-invariant phase of Eqn(6) $\simeq m^2t/E$ evidently
   agrees with the result of Ref.\cite{FritMink} in the non-relativistic
   limit where $E \simeq m$, but such a limit is clearly inappropriate
   to describe experiments with ultra-relativistic neutrinos.
   \par  The most important difference in the treatment given in the 
   present paper to previous ones that have appeared in the literature
   is allowing the possibility for the different neutrino mass eigenstates
   to be produced at different times. Only in this way can the constraints, 
   of both space-time geometry (the detection event is at a unique space-time
    point), and exact energy-momentum conservation in the decay process~\cite{Win},
     be satisfied.
    \par The other new feature is the inclusion of the important
    contribution to the oscillation phase from the propagator of the decaying
    particle, a necessary consequence of the different production times
    of the different mass eigenstates. 
    \par Many recent treatments of the quantum mechanics of neutrino
    oscillations have, following the suggestion of Ref.\cite{Kayser}, used 
    a `wave packet' description of the spatial distributions of the
    source and detector particles. This gives, for the transition 
    probability of the complete production-propagation-detection process: 
  \begin{equation}
 P_{fi} = \left|\sum_m \sum_l\ \sum_{k_1} \sum_{k_2}...\sum_{k_n}\langle \psi_f|f_m\rangle
 \langle f_m| k_1 \rangle
\langle k_1| k_2 \rangle...\langle k_n| i_l \rangle\langle i_l  |\psi_i\rangle \right|^2  
 \end{equation}
    Here, $\psi_i$ and $\psi_f$ are `source' and `detector' wave packets respectively.
  The use of such an equation leads to the prediction, in conflict with Eqn(1) and
  the fundamental law of superposition in quantum mechanics, that {\it different} initial
  or final state particles will interfere with each other~\cite{JHF}. 
   As is well known,
  (for example, in Fermi's `Golden Rules') different initial and final states
   must be (incoherently) summed over at the level of the transition 
   probability, not coherently, at amplitude level, as is done in Eqn(23).
   In a previous covariant analysis of neutrino oscillations~\cite{Moh1},
   the invariant neutrino oscillation phase, derived from Eqn(6) above,
   was found to be multiplied by a factor 1/2 on performing a convolution,
   according to to Eqn(23), with a Gaussian space-time wave packet for the
   source particle. In this way the standard result of Ref.\cite{FritMink}
   was recovered.
\par It may be remarked that the physical interpretation of `neutrino oscillations'
  provided by the path amplitude description is different from the conventional
  one in terms of `flavour eigenstates'. In the latter the amplitudes of different
  flavours in the neutrino are supposed to vary harmonically as a function of time.
  In the amplitudes for the different physical processes in the path amplitude
  treatment there is, instead, no variation of lepton flavour in the propagating
  neutrinos. If the mass eigenstates are represented as superpositions of 
  flavour eigenstates using the inverses of Eqns(2) and (3), there is
   evidently no temporal variation of the lepton flavour composition. 
   Only in the detection process itself are the flavour eigenstates 
   projected out, and the interference effect occurs that is 
   described as `neutrino oscillations'. In the case of the 
   observation of the recoil muons no such projection occurs,
   but exactly similar interference effects are predicted to occur.
   As previously emphasised~\cite{SWS}, the `flavour oscillations'
   of neutrinos, neutral kaons and b-mesons are just special examples
   of the universal phenomenon of quantum mechanical superposition
   that is the physical basis of Eqn(1). 
   \par The most important practical conclusion of the work presented
    here is that identical information on neutrino masses is give by
    the observation of either neutrinos or muons from pion decay at
     rest. In view of the possibility of detecting muons simply and 
    with essentially 100 $\%$ efficiency, in contrast to the tiny
    observable event rates of neutrino interactions, the recent
    indications for $\overline{\nu}_{\mu} \rightarrow \overline{\nu}_e$ 
    oscillations with $\Delta m^2 \simeq $(1eV)$^2$ following $\mu^+$
    decay at rest~\cite{LNSD} could be easily checked by a search 
    for muon oscillations following $\pi^+$ decay at rest. For  
    $\Delta m^2 \simeq $(1eV)$^2$, the first absolute maximum of the
    interference term in Eqn(16) occurs at $L \simeq 8$m. Note that
    Eqn(16) is valid for any muon detection process.  
 
\pagebreak

\newpage
\vspace*{4cm}
\begin{figure}[htbp]
\begin{center}
\hspace*{-0.5cm}\mbox{
\epsfysize15.0cm\epsffile{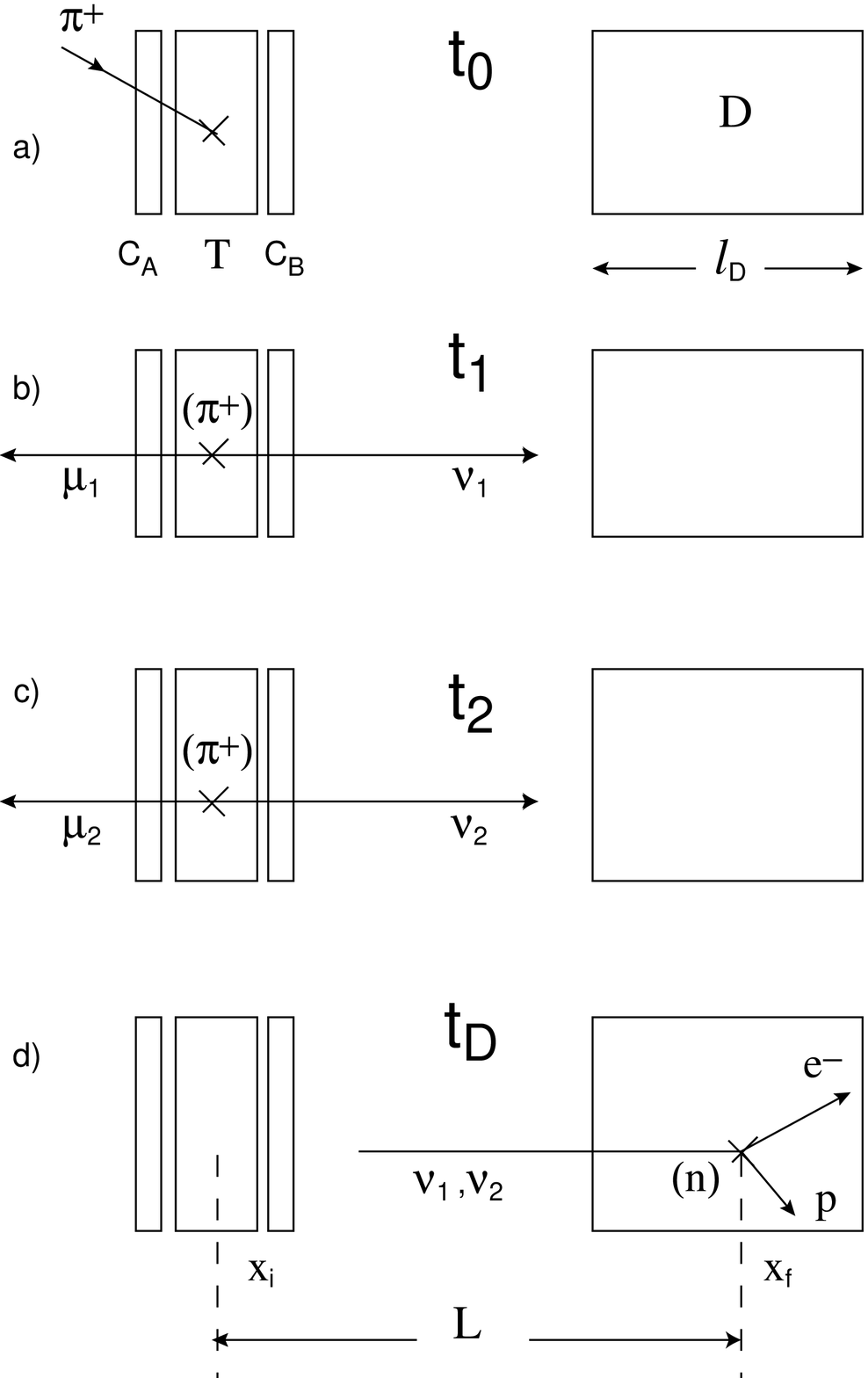}}
\caption{The space-time description of  $\nu_{\mu} \rightarrow \nu_e$ oscillations
  following $\pi^+$ decay at rest, in Feynman's formulation of quantum 
   mechanics (see text).}
\label{fig-fig1}
\end{center}
 \end{figure}
\end{document}